\documentclass{optica-article}

\journal{opticajournal} 

\articletype{Research Article}

\usepackage{lineno}
\usepackage{braket}
\usepackage{hhline}

\begin{document}

\title{All-optical logic gates for extreme ultraviolet switching via attosecond four-wave mixing}

\author{Patrick Rupprecht,\authormark{1,2,*} Daniel M. Neumark,\authormark{1,2} and Stephen R. Leone\authormark{1,2,3}}

\address{\authormark{1}Department of Chemistry, University of California, Berkeley, California 94720, USA\\
\authormark{2}Chemical Sciences Division, Lawrence Berkeley National Laboratory, Berkeley, California 94720, USA\\
\authormark{3}Department of Physics, University of California, Berkeley, California 94720, USA}

\email{\authormark{*}prupprecht@lbl.gov} 


\begin{abstract*} 
All-optical logic-gate-based switching is a prerequisite for photonic computing. This article introduces a logic-gate protocol for noncollinear four-wave mixing (FWM) of one attosecond extreme ultraviolet (XUV) with two few-femtosecond near infrared (NIR) pulses. Simulations show that the NIR carrier-envelope phases (CEPs) alter the spatial distribution of the XUV FWM emission, using doubly-excited states of gas-phase helium as an example. A complete set of logic gates---X(N)OR, (N)AND, and (N)OR---is realized for the 2s3p FWM signal at 63.66\,eV with switching contrasts of 3.6 to 10.4. This theoretical study extends all-optical logic switching to the XUV and x-ray regimes and opens a new pathway for ultrafast photonic logic.

\end{abstract*}

\section{Introduction}
Ultrafast laser pulses are promising candidates to enable petahertz electronics \cite{heide2024petahertz}, which is much anticipated for advancing computation beyond established silicon technology.
Besides boosting clock rates, laser pulses might also help to decrease power consumption by circumventing dissipative electronics altogether and basing computation solely on light \cite{miller2010optical}.
Photonic computing, however, requires switching of light with light, which is more challenging than the interactions of electrons with each other or the interaction of electromagnetic radiation with electrons.
Nonlinear optics is the key to address these challenges and realize all-optical logic gates as basic photonic computing units.
Foremost $\chi^{(3)}$ nonlinearities are used for this purpose, e.g., self-phase modulation \cite{huo2007reconfigurable,wu2020recent,sk2022nonlinear}, cross-phase modulation \cite{jeong1991all,sk2022nonlinear}, and four-wave mixing (FWM) \cite{wang2009experimental,husko2011ultracompact}.
These effects are exploited in microresonators \cite{xu2007all}, optical fibers and waveguides \cite{jeong1991all,huo2007reconfigurable,wang2009experimental,husko2011ultracompact}, interferometric structures \cite{bintjas200020,alquliah2021all}, two-dimensional materials \cite{wu2020recent,sk2022nonlinear}, and quantum dots \cite{dimitriadou2013all,fouskidis2020reconfigurable}. \par
The phase of the electromagnetic field constitutes an efficient control parameter for all-optical logic. 
In the context of ultrafast pulses, this is realized as the carrier-envelope phase (CEP): the phase of the electric field with respect to the pulse envelope.
Measuring and controlling the CEP underlies optical frequency-comb spectroscopy \cite{picque2019frequency} and attosecond science \cite{krausz2009attosecond}.
Hence, experiments with few-cycle pulses that target all-optical switching \cite{schiffrin2013optical}, logic \cite{boolakee2022light}, and data encoding \cite{hui2023ultrafast} often rely on the CEP of the input pulses as the control parameter.\par
The next step after photonic logic in the infrared spectral regime is a shift towards higher carrier frequencies and hence towards the extreme ultraviolet (XUV; from approximately $10-100$\,eV) region.
While one optical cycle at 1550\,nm is about 5.2\,fs, the cycle time of XUV light at 60\,eV ($\lambda_{XUV}=20.6$\,nm) is only 69\,as.
With the rise of attosecond technology, producing and controlling attosecond XUV pulses has become a routine operation via high-order harmonic generation (HHG) \cite{corkum1993plasma}.
Realizing logic gates that involve XUV light, however, imposes challenges:
As solid-state materials have a large absorption cross-section in the XUV, noble gases are the target of choice to mediate a logic operation.
Furthermore, due to the low efficiency of the HHG process, the resulting XUV pulses are in general too weak to cause a nonlinear response themselves.
Hence, as a first step towards XUV-based logic applications, a scheme in which XUV pulses can be switched by CEP-manipulations of near-infrared (NIR) pulses is desirable.\par
In this study, noncollinear FWM of two few-cycle NIR pulses with one attosecond XUV pulse in a gas-phase sample is proposed as a pathway towards all-optical logic with XUV light.
An experimentally readily accessible scenario that applies attosecond FWM to doubly-excited states in helium near 60\,eV is simulated.
The CEP of the NIR input pulses is used as logic input basis, while the spatially-isolated XUV FWM emission acts as boolean output. 
In fact, a complete set of logic gates and their inversions can be realized by changing the CEP input basis accordingly. 

\section{Noncollinear attosecond four-wave-mixing scheme}

The proposed experimental noncollinear attosecond FWM setup for logic-gate-based switching of XUV light is schematically depicted in Fig.~1: 
\begin{figure}[htbp]
\centering\includegraphics[width=\linewidth]{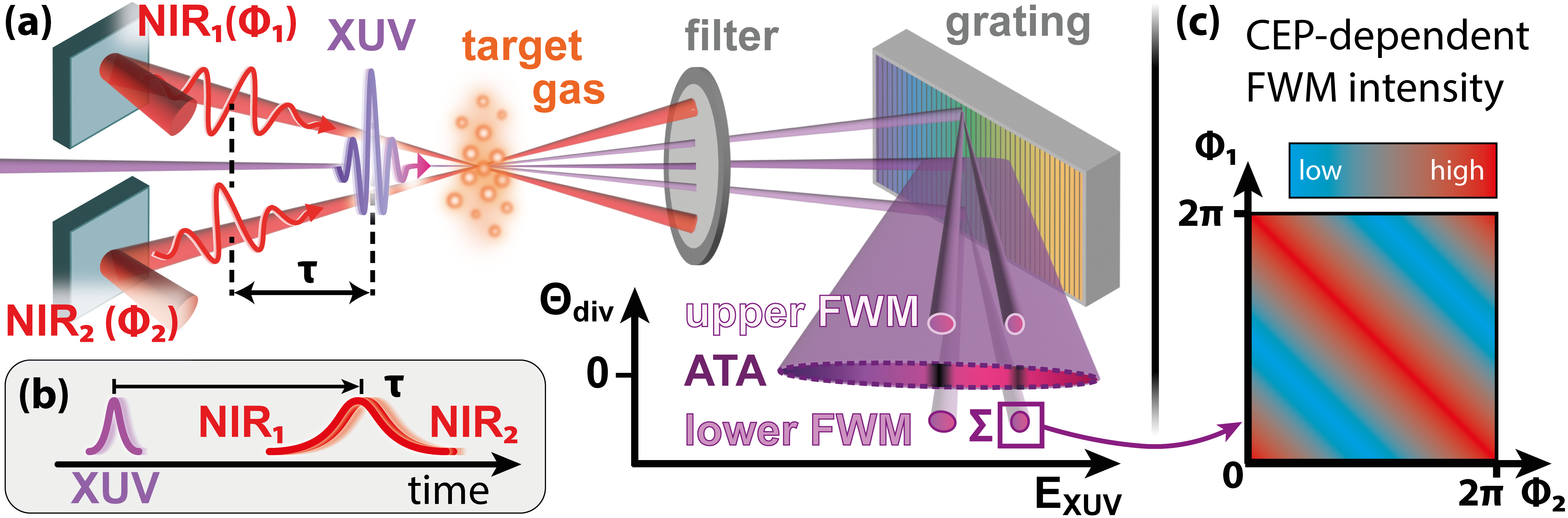}
\caption{Attosecond four-wave-mixing scheme. (a) Experimental setup for attosecond FWM logic switching. An attosecond XUV pulse is focused together with two time-delayed ($\tau$) few-cycle pulses NIR$_{1,2}$ into a gas-phase target. Afterwards, a thin metal filter cleans the XUV signals from the residual NIR pulses. Due to the noncollinear angle of the NIR beams with respect to the XUV beam, the resulting FWM beams are spatially separated from the on-axis attosecond transient absorption (ATA) beam. A grating  spectrally disperses the XUV beams, e.g., the FWM signal into its Rydberg series. (b) Pulse sequence. The XUV pulse precedes the NIR pulse pair by $\tau$. (c) CEP-dependent FWM intensity. The measured FWM intensity of one resonance is binned and measured with respect to its dependence on the NIR CEPs $\Phi_{1,2}$.}
\end{figure}
This technique mixes one XUV photon with two NIR photons in a third-order nonlinear process. 
Due to noncollinear angles between the XUV beam and the two NIR beams, the resulting degenerate XUV FWM signals are emitted at an angle with respect to the transmitted attosecond transient absorption (ATA) beam.
Details on the phase-matching relation that leads to the FWM emission angle are given in the Supplemental Document (SD) section~1.
If the NIR noncollinear input angle $\Theta$ is chosen large enough, the FWM emission angle $\Theta_{FWM}$ can exceed the natural divergence of the ATA beam, which results in spatially isolated, background-free FWM signals.
The ideal target to mediate the nonlinear process should be mostly transparent to XUV light, yet exhibit distinct XUV-excited states, which are directly accessible from the electronic ground state (so-called \textit{bright} states).
Moreover, further highly excited states (so-called \textit{dark} states) should exist, which can be dipole-coupled by the NIR pulses from the XUV-excited bright states but not from the ground state.
Such an electronic structure guarantees a resonant enhancement of the FWM process and is often the case for noble gases, where short-lived XUV-excited states are embedded in the continuum above the first ionization threshold.
After the FWM process, the residual NIR light is removed by a thin metal filter (e.g., aluminum), which is partially transparent for XUV light but not for the NIR.
The transmitted XUV beams are then spectrally dispersed by a reflective grating and their spectra are measured with an XUV-sensitive charge-coupled device (CCD) camera chip. 
Here, the FWM signals can be separated from the ATA signal by the divergence angle, while FWM signals involving different bright states are deciphered by their emission photon energies. \par
So far, noncollinear attosecond FWM has been successfully used for measurements of bright- and dark-state dynamics by scanning the temporal delay between the XUV and both NIR pulses or the XUV/NIR$_1$ pulses and the NIR$_2$ pulse, respectively \cite{cao2016noncollinear,cao2018excited,marroux2018multidimensional,fidler2019autoionization,fidler2019nonlinear,fidler2020self,lin2021coupled,gaynor2021solid,fidler2022state,gaynor2023nonresonant,puskar2023measuring,rupprecht2024extracting,yanez2025nonperturbative,puskar2025probing}.
The setup shown in Fig.~1 was used for the FWM experiments conducted at Berkeley, but without scanning the CEP of the NIR pulses.
For ultrafast logic and petahertz electronics, though, the CEP is the control parameter of choice as it can be rapidly changed in an accurate manner and this change acts intrinsically on the (sub-)optical cycle timescale \cite{ritzkowsky2024chip}. 
Therefore, we propose a FWM scheme in which the CEPs $\Phi_{1,2}$ of the NIR pulses are scanned instead of the time delays between the laser pulses.
The fixed pulse sequence as shown in Fig.~1(b) has a small time delay $\tau$ between the XUV and both temporally overlapped NIR pulses.
If $\tau$ is chosen to be larger than the NIR pulse duration but shorter than the lifetimes of the involved XUV bright states, this sequence guarantees robustness against small XUV-NIR time-delay variations.
This is experimentally relevant as the XUV and the NIR beams normally form a Mach-Zehnder interferometer with multi-meter leg length. 
Overall, the intensity of one FWM signature will be binned and represented in dependence of $\Phi_{1,2}$ in a two-dimensional measurement trace as illustrated in Fig.~1(c).
Hence, this study explores conditions under which the CEP-dependent FWM signals can be observed and how this CEP-dependence can be exploited for all-optical logic gates acting on the XUV FWM emission.

\section{Few-level four-wave-mixing simulation}

While the FWM intensity evolution of time-delay scans can be intuitively linked to quantum-state lifetimes for perturbative NIR intensities, this intuition breaks down when strong-field effects are involved.
As such strong-field effects are crucial for achieving a maximum XUV switching contrast (see section~4.2.), a simulation is required to predict the CEP-dependent and divergence-resolved FWM intensity distribution accurately.\par
A few-level time-dependent Schrödinger equation (TDSE) calculation is combined with a multi-emitter model similar to an approach by Mi \textit{et al.} \cite{mi2021method} for simulating the measurement observable, which is the FWM spectrogram.
This simulation scheme is shown in Fig.~2.\par
\begin{figure}[htbp]
\centering\includegraphics[width=\linewidth]{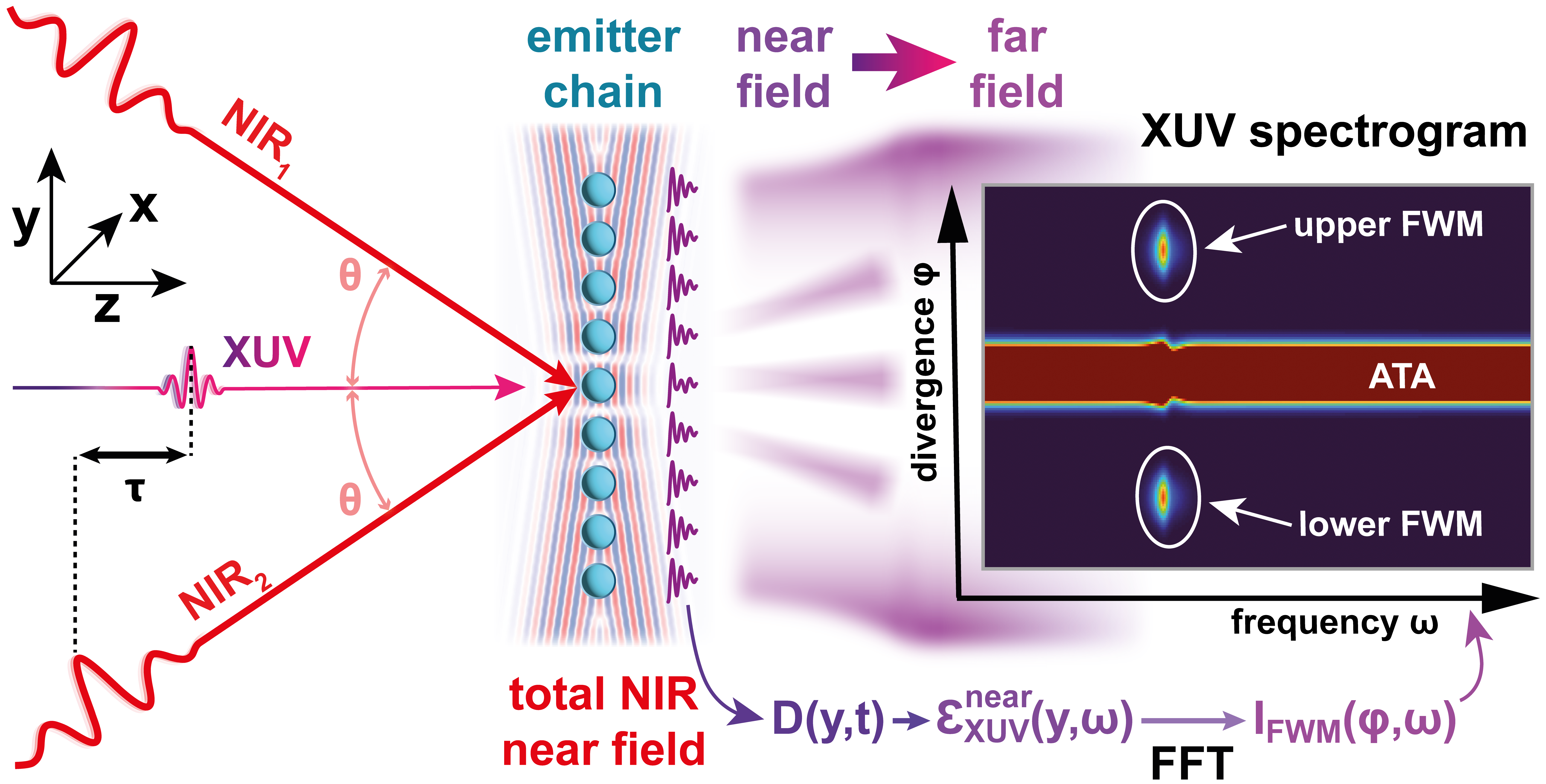}
\caption{Four-wave-mixing simulation scheme. The noncollinear angle ($2\Theta$) between the NIR$_{1,2}$ beams results in a periodically modulated (along the y-axis) electric-field superposition in the target volume. The gas-phase target is approximated as a one-dimensional emitter chain along the y direction. Solving the TDSE for the input XUV field in combination with the space- and time-dependent NIR field superposition results in a time-dependent dipole moment $D(y,t)$ that differs on the 10\,µm-scale of the NIR field modulations. The resulting electric near-field signal $\mathcal{E}^{near}_{XUV}(y,\omega)$ is propagated via a Fourier transform (FFT) into the far field. The far-field divergence-dependent XUV-intensity spectrogram $I_{FWM}(\varphi,\omega)$ is the measurement observable in Fig.~1.}
\end{figure}
In the simulation, the Hamiltonian $\mathcal{H}$ of each emitter can be separated into the time-dependent Hamiltonians $\mathcal{H}_{NIR}(t)$ and $\mathcal{H}_{XUV}(t)$ of the light-matter interactions, as well as the time-independent Hamiltonian $\mathcal{H}_0$, which captures the evolution of the quantum states:
\begin{equation}
    \mathcal{H}(t) = \mathcal{H}_{XUV}(t) + \mathcal{H}_{NIR}(t) + \mathcal{H}_0 \ .
\end{equation}
While the separation of $\mathcal{H}_{NIR}(t)$ and $\mathcal{H}_{XUV}(t)$ is not strictly necessary, it prevents any numerical artifacts in the specific case of a weak XUV pulse and intense NIR pulses.
The respective Hamiltonians are given in the SD section~2 for the scenario of two bright states and one dark state.\par
The TDSE is solved on a discrete time grid with time steps $\Delta t$ in a split-step manner as discussed by Bandrauk \& Shen \cite{bandrauk1993exponential}, where $\mathcal{H}_{0}$ is propagated for half a time step $\Delta t/2$.
Afterwards, the time-dependent Hamiltonians $\mathcal{H}_{XUV/NIR}$ are rotated into their time-dependent diagonal basis and propagated for a full time step $\Delta t$.
Finally, $\mathcal{H}_{0}$ is propagated for another half time step $\Delta t/2$.
While the initial diagonalization of the time-independent Hamiltonian $\mathcal{H}_{0}$ is valid on the whole time grid, the time-dependent Hamiltonians $\mathcal{H}_{XUV/NIR}$ have to be diagonalized for each time step in principle.
As the time grid has to be fine enough to sample the frequency of the emitted XUV light, and at the same time the length of the temporal grid determines the resulting spectral resolution, the temporal grid size tends to be on order of $10^5$.
To reduce the computational cost, the XUV and NIR electric fields are precalculated for the whole emitter chain in a field-quantized manner. 
This allows to efficiently calculate a reduced number of time-dependent Hamiltonians in their diagonal bases before conducting the split-step temporal propagation.\par
The spatially isolated FWM signals arise from constructive interference of an emitter ensemble.
The isotropic, low-density gas-phase target is approximated by a one-dimensional chain of equidistant emitters in the plane of all three pulses and orthogonal to the XUV beam (along the y-axis in Fig.~2).
Then, the spatially dependent electric fields along the emitter chain act as the input for the TDSE of each emitter.
The resulting wavefunctions $\psi_i(y,t)$ for every quantum state $i$ provide the populations $|\psi_i|^2(y,t)$ and the time-dependent dipole moment
\begin{equation}
    D(y,t) = \sum_{i\in \{b_1, d, b_2\}}\bra{\psi_g(y,t)}\hat{d}\ket{\psi_i(y,t)} \ .
\end{equation}
Here, $\hat{d}$ denotes the dipole operator, which is only non-zero for bright-to-ground state transitions.
$D(y,t)$ is directly linked to the complex absorbance $\widetilde{OD}(y,t)$ of each emitter via the absorption cross section \cite{gaarde2011transient}, and hence the complex-valued optical near-field signal $\mathcal{E}^{near}_{XUV}$ is given by a Fourier transform $\mathcal{F}$ from the time into the frequency domain:
\begin{equation}
    \mathcal{E}^{near}_{XUV}(y,\omega) = 10^{-\widetilde{OD}(y,t)} \ \ \text{with} \ \ \widetilde{OD}(y,\omega) \propto \omega \frac{\mathcal{F}\left(D(t)\right)(y)}{\mathcal{E}_{XUV}(\omega)} \ .
\end{equation}
Finally, the measured XUV spectrogram $I_{FWM}(\varphi,\omega) = \left|\mathcal{E}^{far}_{XUV}\right|^2(\varphi, \omega)$ is given by propagating $\mathcal{E}^{near}_{XUV}(y,\omega) $ via Fraunhofer-diffraction and hence spatial Fourier transform into the far field. \par
This simulation has been recently applied by the authors to evaluate FWM signatures of $4d^{-1}6\ell$ core-level excited states of gas-phase xenon under the influence of NIR-induced strong-field effects \cite{puskar2025probing}.
Besides this multi-emitter TDSE approach, which is conducted on an explicit spatial grid, an alternative, more analytical FWM simulation by Yanes-Pagans \textit{et al.} enabled the investigation of strong-field associated FWM signals for $3s^{-1}$-excited states in argon \cite{yanez2025nonperturbative}.

\section{Results for doubly-excited states in helium}

The multi-emitter few-level TDSE simulation was applied to gas-phase helium as a model system for XUV logic-gate-based switching.
States in which both electrons of the helium atom are excited by one XUV photon at the same time, so-called doubly-excited states, are ideal candidates for a proof-of-principle study:
The lowest-lying doubly-excited states in helium are located around 60\,eV and hence are readily accessible with attosecond XUV pulses originating from HHG \cite{calegari2016advances}.
These states have been extensively theoretically investigated as a showcase scenario for electronic correlation \cite{fano1983correlations,rost1997resonance,tanner2000theory}.
Furthermore, the lowest-lying doubly-excited states in helium are embedded in the $1s$ ionization continuum and are therefore metastable with lifetimes in the 10s to 100s of femtoseconds \cite{domke1996high,gilbertson2010monitoring,rupprecht2024extracting}.
The autoionizing nature of these states is also imprinted in the Fano character of their spectral absorption line shapes \cite{fano1961effects}.
These characteristics make doubly-excited states in helium a popular system for experiments targeting ultrafast control of correlated quantum dynamics \cite{gilbertson2010monitoring,ott2013lorentz,ott2019strong}.
Recently, Mi \textit{et al.} \cite{mi2021method} theoretically studied an extraction of doubly-excited-state lifetimes in helium with noncollinear attosecond FWM in the perturbative regime, while Rupprecht \& Puskar \textit{et al.} \cite{rupprecht2024extracting} experimentally applied attosecond FWM to systematically measure the 2s$n$p Rydberg-series bright-state and the $2p^2$ dark-state lifetimes with a FWM setup as depicted in Fig.~1(a), but without CEP stabilization.

\subsection{Few-level helium model}
The ground state and three doubly-excited states of helium [see Fig.~3(a)] are chosen as a basis set for the few-level FWM simulation in helium.
\begin{figure}[htbp]
\centering\includegraphics[width=\linewidth]{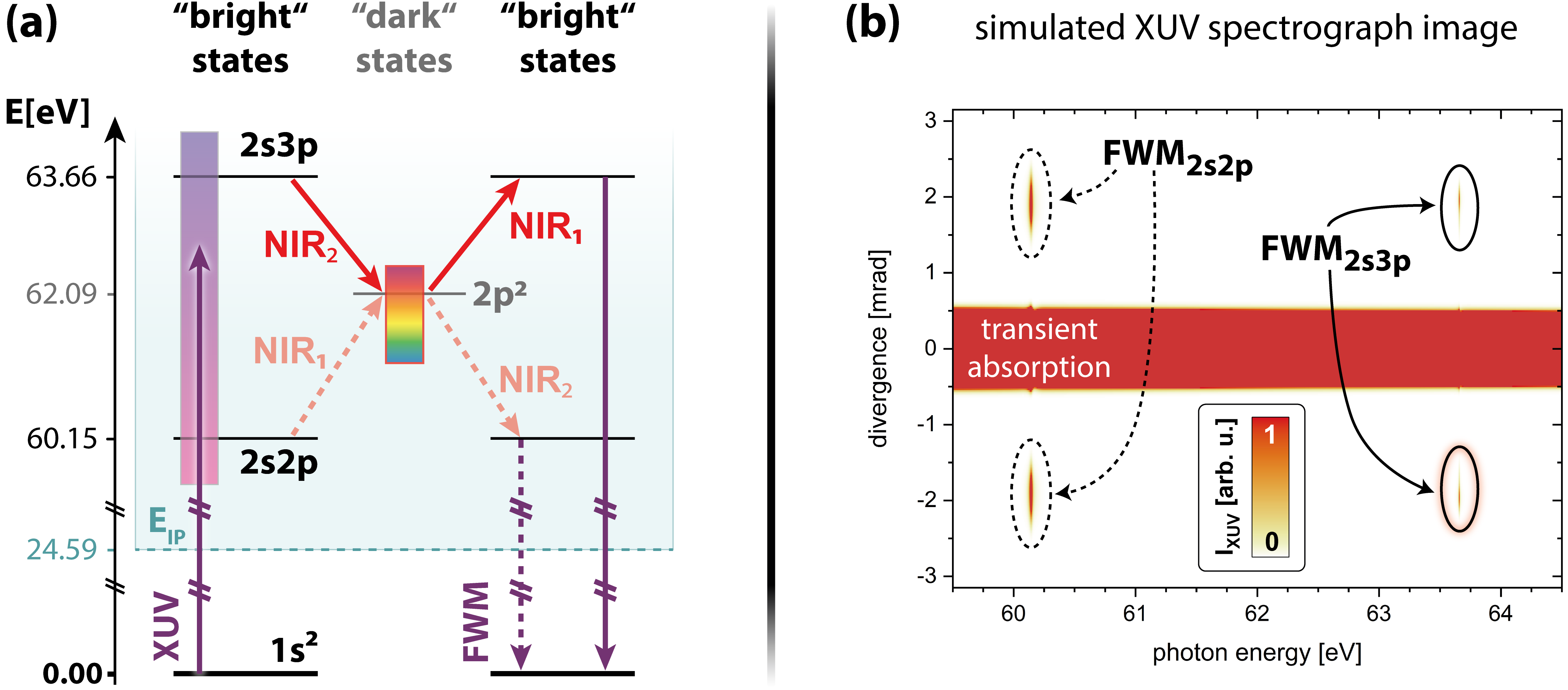}
\caption{FWM simulation applied to doubly-excited states in helium. (a) Few-level model scheme of helium. A broadband attosecond XUV pulse populates the bright states 2s2p and 2s3p, located above the first ionization  threshold ($E_{IP}$). The NIR$_{1,2}$ pulses lead to FWM emissions by a $\Lambda$- (dashed orange) or $V$-coupling (solid red) scheme via the 2p$^2$ dark state. The rainbow-colored bar indicates the spectral bandwidth of the NIR pulses. (b) FWM simulation for a noncollinear input angle of $\Theta = 2^{\circ}$, an intensity of $I(NIR_{1,2}) = 1 \times 10^{10}$\,W/cm$^{2}$, CEP values of $\Phi_{1,2} = 0$, and a time delay of $\tau = 10$\,fs. In the following, the focus is on the highlighted, lower-right 2s3p FWM signal.}
\end{figure}
Two members of the 2s$n$p Rydberg series act as bright states: the 2s2p and the 2s3p states, located at 60.15\,eV and 63.66\,eV, respectively.
Furthermore, the 2p$^2$ dark state is included, which is located at 62.09\,eV.
Utilizing 5\,fs few-cycle NIR pulses centered around 800\,nm results in resonant coupling pathways from either bright state.
The transition dipole matrix elements are given in the literature by $d_{2s2p-2p^2} = 2.17$\,a.u. (atomic units) \cite{loh2008femtosecond} and $d_{2s3p-2p^2} = 0.81$\,a.u. \cite{ott2014reconstruction}, large values which are crucial for the strong-field effects that boost the switching contrast.
Further TDSE parameters are given in the SD section~3.
An exemplary simulated spectrogram is shown in Fig.~3(b) in the weak-field limit of $I(NIR_{1,2}) = 1.0 \times 10^{10}$\,W/cm$^2$ and for NIR CEP values of $\Phi_{1,2} = 0$.
In the following, we will focus on the lower FWM signature at the 2s3p energy of 63.66\,eV [highlighted in Fig.~3(b)], which originates from the 2s3p$\leftrightarrow$2p$^2$ $V$-coupling pathway [solid red arrows in Fig.~3(a)].

\subsection{Strong-field effects}
A more detailed depiction of the 2s3p-associated lower FWM signature [compare Fig.~3(b)] at perturbative NIR intensities of $10^{10}$\,W/cm$^2$ is shown in Fig.~4(a) and (b) for CEP values of $\Phi_{1,2} = \pi/2$ and $\Phi_{1,2} = 0$, respectively.
\begin{figure}[htbp]
\centering\includegraphics[width=\linewidth]{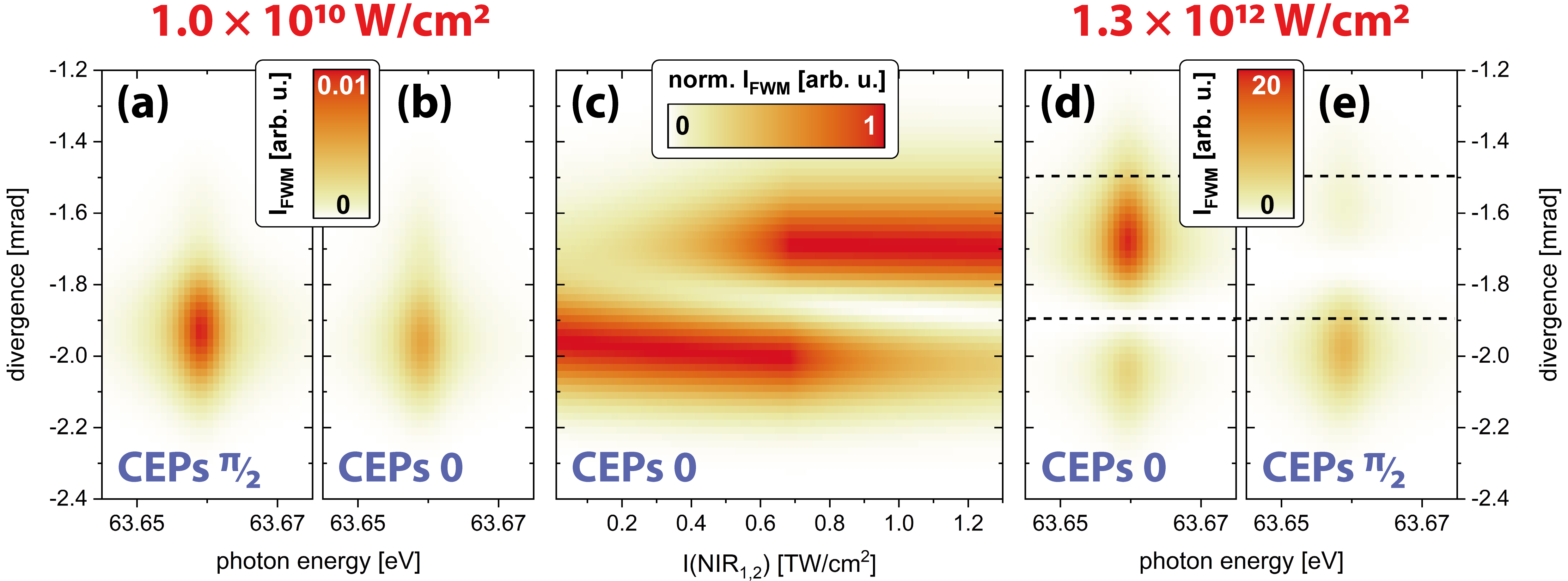}
\caption{Strong-field effects on the helium 2s3p lower FWM emission. Simulated FWM emission in the low-intensity [$I(NIR_{1,2}) = 1.0 \times 10^{10}$\,W/cm$^{2}$] limit for CEPs (a) $\Phi_{1,2} = \pi/2$ and (b) $\Phi_{1,2} = 0$. (c) NIR$_{1,2}$-intensity dependence of the photon-energy binned FWM divergence distribution for $\Phi_{1,2} = 0$. FWM emission in the high-intensity [$I(NIR_{1,2}) = 1.3 \times 10^{12}$\,W/cm$^{2}$] regime for (d) $\Phi_{1,2} = 0$ and (e) $\Phi_{1,2} = \pi/2$. The dashed black lines in (d) and (e) indicate the divergence region of interest.}
\end{figure}
This comparison reveals that a CEP variation can only moderately modify the respective CEP intensity by about a factor of two.\par
However, if the strong-field regime is accessed by increasing the NIR intensities up to $1.3\times 10^{12}$\,W/cm$^2$, this situation changes dramatically. 
Firstly, when keeping the CEPs fixed ($\Phi_{1,2} = 0$), the FWM intensity distribution along the divergence axis changes towards a double-peak structure as depicted in the Fig.~4(c) NIR$_{1,2}$-intensity scan.
Secondly, a direct comparison of the FWM spectrograms at these higher NIR intensities for CEPs $\Phi_{1,2} = 0$ and $\Phi_{1,2} = \pi/2$ [Figs.~4(d) and (e)] illustrates a much stronger dependence of the FWM intensity and its divergence distribution on the CEP compared to the low NIR$_{1,2}$-intensity case.
In the upper divergence region between -1.9 and -1.5\,mrad [indicated by the black dashed lines in Fig.~4(d) and (e)], the FWM intensity nearly vanishes at $\Phi_{1,2} = \pi/2$ in comparison to its distinct maximum at $\Phi_{1,2} = 0$.
Therefore, this divergence region-of-interest (ROI) is suitable for relatively high-contrast CEP-dependent XUV switching.
While a further NIR-intensity increase might enhance the CEP-dependent contrast, the NIR intensity is kept at a moderate level of $1.3\times 10^{12}$\,W/cm$^2$ for this theoretical study to ensure the validity of the utilized four-level basis set, as otherwise nonresonant couplings to further Rydberg states might not be negligible anymore.\par
Some intuitive insights into the impact of strong-field effects on the FWM signatures are provided by the population dynamics of an exemplary emitter in the middle of the emitter chain and on-axis to the incoming XUV beam (compare Fig.~2).
Figs.~5(a) and (b) show the population comparison for the weak- vs. strong-field case for CEPs $\Phi_{1,2} = 0$ and hence for the central emitter experiencing the highest combined NIR field strength.
\begin{figure}[htbp]
\centering\includegraphics[width=\linewidth]{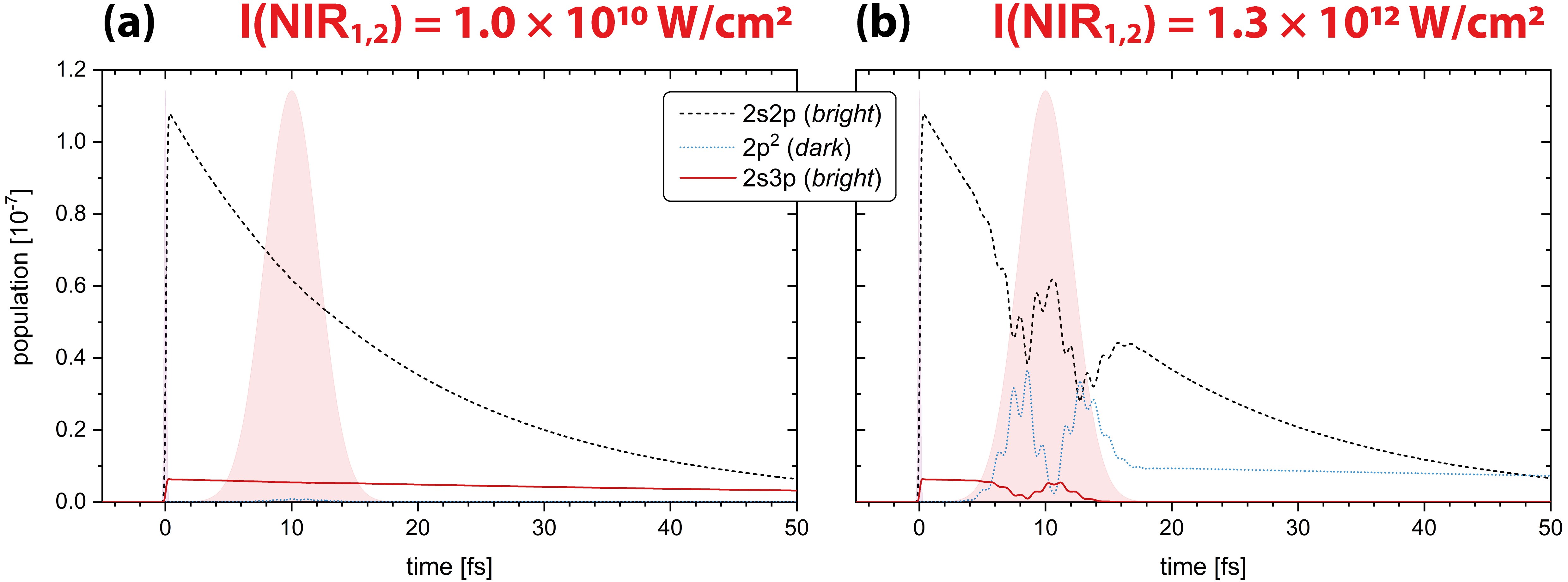}
\caption{Simulated population evolution for one central on-axis emitter in the (a) low- [$I(NIR_{1,2}) = 1.0 \times 10^{10}$\,W/cm$^{2}$] and (b) high-intensity [$I(NIR_{1,2}) = 1.3 \times 10^{12}$\,W/cm$^{2}$] regimes. The electric-field envelope of the NIR pulses is given as shaded red area while the attosecond XUV pulse is centered around 0\,fs.}
\end{figure}
The exponential decays of the 2s2p (black dashed line) and 2s3p (red solid line) bright states remain mainly unperturbed by the NIR pulses and decay according to their natural lifetimes in Fig.~5(a) in the case of low NIR intensities ($1.0\times 10^{10}$\,W/cm$^2$).
Here, the 2p$^2$ dark state (dotted blue line) is barely populated and can be hardly seen.
For higher NIR intensities of $1.3\times 10^{12}$\,W/cm$^2$, however, considerable Rabi cycling occurs in Fig.~5(b).
While the Rabi cycling of the 2s2p state never totally depletes the state's population and ends up repopulating it within the NIR pulse envelopes, the 2s3p state is completely quenched after the NIR pulses. 
In consequence, the periodic NIR field structure does not proportionally imprint onto the time-dependent dipole moment and the near-field emission for the strong-field case.
Instead, regions of higher NIR intensities within the transient NIR field structure lead to a 2s3p population depletion and therefore a distortion of the near-field signal periodicity and spatial frequencies.
The distortion of spatial frequencies of the near-field emission translates via Fourier transform (FFT in Fig.~2) to a modified FWM divergence distribution in the far field as depicted in Fig.~4(c). 
Altering the CEPs of few-cycle NIR pulses results in a different achievable peak intensity and a shift of the transient periodic field pattern within the spatio-temporal NIR pulse overlap in the target.
In the weak-field case with its linear relationship between the NIR intensity and the excited-state population, this leads to proportional alterations in the FWM intensity.
The strong-field scenario, however, leverages the nonlinear population behavior with respect to the NIR intensity pattern due to Rabi cycling to achieve much larger differences in the FWM intensity and divergence structure for the same CEP changes.

\subsection{Logic-gate implementations}
In order to exploit this strong-field-induced CEP dependence of the FWM intensity and divergence characteristics, a systematic scan of both CEPs varied independently at NIR intensities of $1.3 \times 10^{12}$\,W/cm$^2$ was simulated as shown in Fig.~6.
\begin{figure}[htbp]
\centering\includegraphics[width=\linewidth]{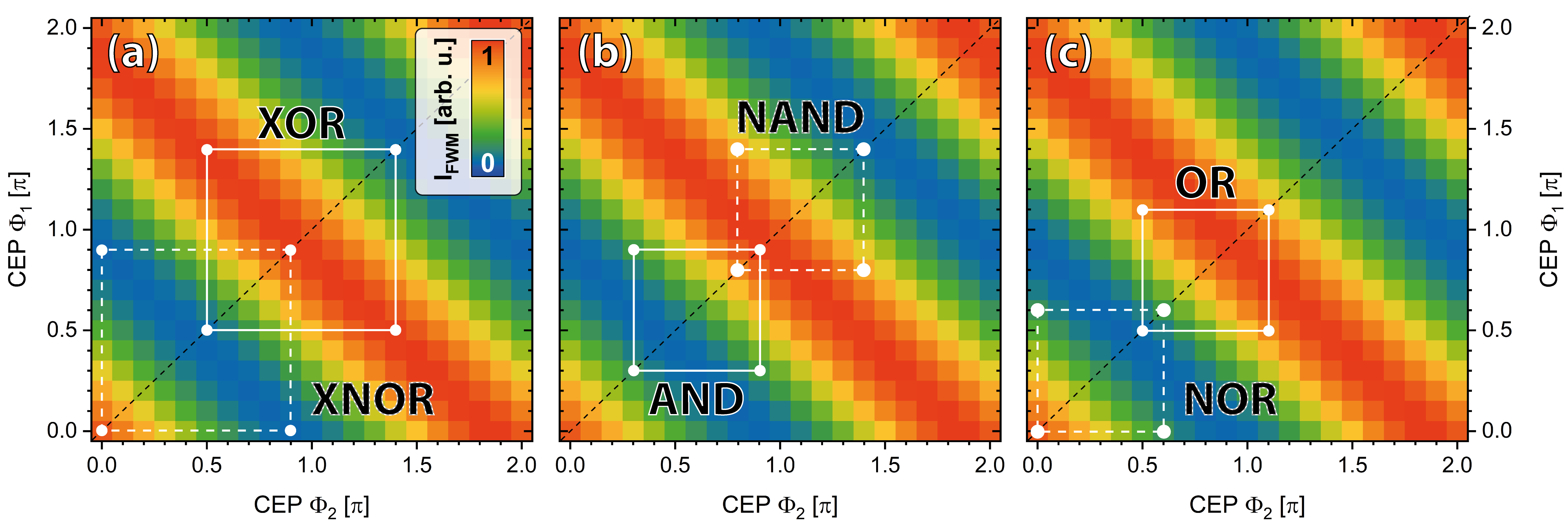}
\caption{CEP-dependence of the 2s3p FWM signal within the $-1.9$ to $-1.5$\,mrad divergence region at $I(NIR_{1,2}) = 1.3 \times 10^{12}$\,W/cm$^{2}$. (a) CEP basis set of $(0.5, 1.4)\pi$ and $(0.0, 0.9)\pi$ for X(N)OR gates, (b) basis set of $(0.3, 0.9)\pi$ and $(0.8, 1.4)\pi$ for (N)AND gates, and (c) basis set of $(0.5, 1.1)\pi$ and $(0.0, 0.6)\pi$ for (N)OR gates. The specific relative FWM intensities and logic-gate contrasts are summarized in Table~1.}
\end{figure}
Each pixel of the resulting two-dimensional map corresponds to the 2s3p FWM intensity binned over the photon energy of the resonance and the divergence ROI (-1.9 to -1.5\,mrad).
Both CEPs are scanned from 0 to $2\pi$ in $0.1\pi$ steps and the trace is normalized to the highest pixel value.\par
Several different logic gates are considered. 
For each logic gate, an optimal CEP input basis is chosen that allows for a maximum contrast under the respective logic operation.
In the nomenclature used, the lower CEP value represents a logic 0 of the input basis, while the higher value equals a logic 1.
In a similar manner, a low FWM value is identified with a logic 0 as output, while a higher FWM value represents a logic 1.
Fig.~6(a) shows the first pair of logic gate implementations: \textit{exclusive (not) or} gates [X(N)OR].
An XOR gate is realized for the CEP input basis set $\Phi_{1,2} \in \{0.5, 1.4\}$ [solid white square in Fig.~6(a)].
If both CEP input values are the same ($\Phi_1 = \Phi_2$), a very low FWM value of $\sim 0.1$ is achieved, while in the other two cases, the FWM intensity is close to 1.
The inverse XNOR gate can be implemented by a $-0.5\pi$ offset to the basis set [dashed white square in Fig.~6(a)].
In general, a gate and its inverse share the same $\Delta\Phi$, but their input basis sets are shifted along the $\Phi_1 = \Phi_2$ diagonal of the plot (black dashed line in Fig.~6).
This is rooted in the symmetry of the CEP map along its diagonal.\par
To further investigate systematics in the choice of suitable CEP basis sets for logic-gate design, high-resolution on-axis and diagonal lineouts of Fig.~6 are depicted in Figs.~7(a) and (b), respectively.
\begin{figure}[htbp]
\centering\includegraphics[width=\linewidth]{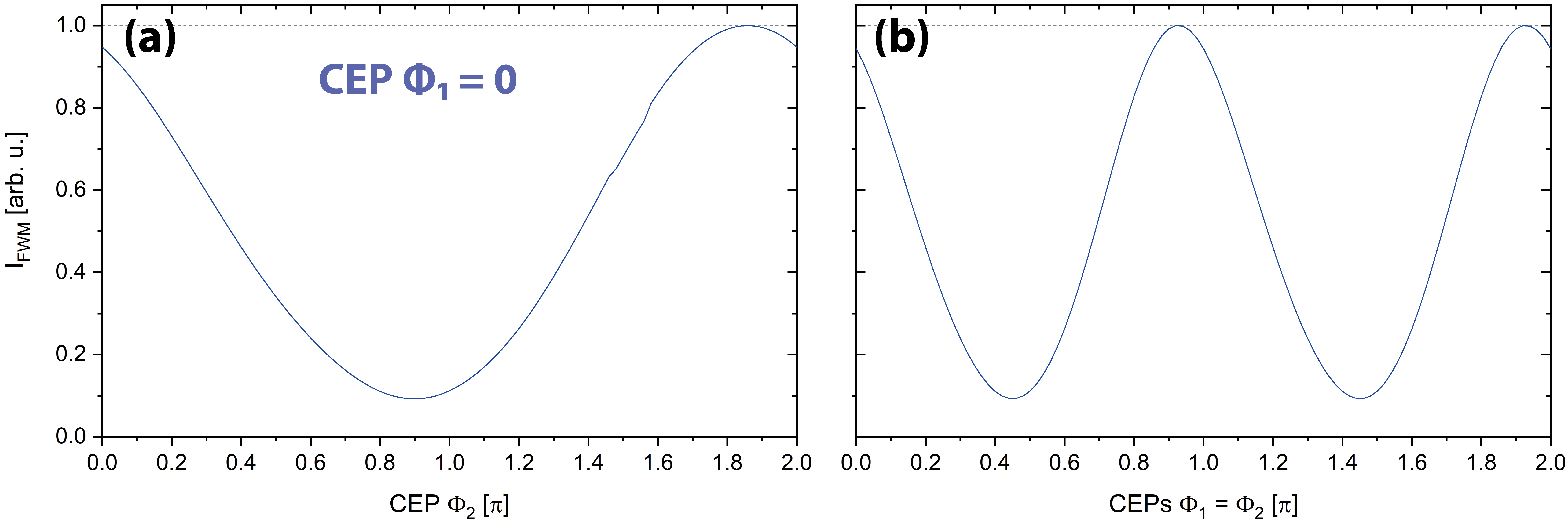}
\caption{Dependence of the FWM intensity on the CEP when varying (a) one CEP ($\Phi_2$; $\Phi_1 = 0$) or (b) both CEPs at the same time. Here, (a) represents the lineout along the x-axis of Fig.~6, while (b) is the lineout along the diagonal (black dashed line in Fig.~6).}
\end{figure}
The lineout along the $\Phi_2$ axis depicted in Fig.~7(a) shows an oscillatory behavior with one minimum and maximum, while the diagonal lineout shown in Fig.~7(b) has two minima and maxima. 
As the input logic points $(0, 0)$ and $(1, 1)$ are located along the diagonal, Fig.~7(b) constrains suitable CEP values for the respective logic gate. 
In a second step, a compromise is found with the on-axis lineout plot Fig.~7(a) that has to be phase-shifted if the fixed CEP is not 0.\par
Hence, the chosen $\Delta\Phi = 0.9\pi$ for the X(N)OR logic gates could be directly deduced from the diagonal lineout plot.
The other basic logic gates, namely the \textit{(not) and} gates [(N)AND] and the \textit{(not) or} gates [(N)OR], though, have different diagonal values and therefore their basis sets show a significantly reduced $\Delta \Phi = 0.6$ value.
Due to the limited resolution of the computationally costly 2D maps in Fig.~6, the logic gate basis sets presented here vary by up to $\Delta \Phi = \pm 0.1\pi$ from the ideal extracted values of the high-resolution lineouts in Fig.~7.
A summary of the logic gate implementations, their basis sets, and their minimum switching contrasts is given in Table~1.
\begin{table}[htbp]
	\caption{Summary of logic gate parameters. The respective logic gate implementations are depicted in Fig.~6. Logic values are given in bold numbers \textbf{0} and \textbf{1} and in brackets after the FWM values in the I$_{FWM}$ columns. The minimum contrast values are given as the ratio between the highest I$_{FWM}$ value for logic \textbf{0} and the smallest I$_{FWM}$ value for logic \textbf{1} of the respective gate implementation.}
	\label{tab:example2}
 \centering
\begin{tabularx}{1\textwidth}{ >{\centering\arraybackslash}X | >{\centering\arraybackslash}X | >{\centering\arraybackslash}X | >{\centering\arraybackslash}X | >{\centering\arraybackslash}X | >{\centering\arraybackslash}X | >{\centering\arraybackslash}X}
\hline
gate & \multirow{2}{*}{\shortstack{CEP input \\basis set\vspace{2pt}}} & \multicolumn{4}{ c |}{normalized I$_{FWM}$ from Fig.~6 for} & \multirow{2}{*}{\shortstack{minimum \\contrast\vspace{5pt}}}\\
\cline{3-6}
&&(\textbf{0}, \textbf{0})&(\textbf{0}, \textbf{1})&(\textbf{1}, \textbf{0})&(\textbf{1}, \textbf{1})\\
\hhline{=|=|=|=|=|=|=}
XOR & $(0.5, 1.4)\pi$ & 0.11 \ [\textbf{0}] & 0.99 \ [\textbf{1}] & 1.00 \ [\textbf{1}] & 0.11 \ [\textbf{0}] & 9.0\\
XNOR & $(0.0, 0.9)\pi$ & 0.94 \ [\textbf{1}] & 0.09 \ [\textbf{0}] & 0.09 \ [\textbf{0}] & 0.99 \ [\textbf{1}] & 10.4\\
\hline
AND & $(0.3, 0.9)\pi$ & 0.24 \ [\textbf{0}] & 0.26 \ [\textbf{0}] & 0.26 \ [\textbf{0}] & 0.99 \ [\textbf{1}] & 3.8\\
NAND & $(0.8, 1.4)\pi$ & 0.83 \ [\textbf{1}] & 0.73 \ [\textbf{1}] & 0.73 \ [\textbf{1}] & 0.11 \ [\textbf{0}] & 6.6\\
\hline
OR & $(0.5, 1.1)\pi$ & 0.11 \ [\textbf{0}] & 0.83 \ [\textbf{1}] & 0.83 \ [\textbf{1}] & 0.73 \ [\textbf{1}] & 6.6\\
NOR & $(0.0, 1.6)\pi$ & 0.94 \ [\textbf{1}] & 0.24 \ [\textbf{0}] & 0.24 \ [\textbf{0}] & 0.26 \ [\textbf{0}] & 3.6\\
\hline
\end{tabularx}
\end{table}

\section{Conclusion}
This article develops a framework for all-optical X(N)OR, (N)AND, and (N)OR logic gates for switching phase-matched XUV emission in table-top attosecond four-wave mixing via CEP control of NIR few-cycle pulses.
An envisioned noncollinear XUV/NIR FWM experiment on doubly-excited states in helium builds on earlier CEP-averaged FWM work of the authors.
The CEP-dependent experiment is simulated with a newly developed multi-emitter few-level TDSE code by including the ground state, the 2s2p and 2s3p bright states as well as the 2p$^2$ dark state in the $60-65$\,eV XUV spectral region.
Future XUV-switching applications rely on a high switching contrast.
Therefore, strong-field effects are leveraged for achieving switching contrasts in the range of 3.6 to 10.4 that would otherwise remain $\leq 2$ in the perturbative regime.
At NIR intensities of $1.3 \times 10^{12}$\,W/cm$^2$, considerable Rabi cycling leads to a significant CEP-dependence of the emitted 2s3p FWM signature and its divergence distribution.\par
This study demonstrates the feasibility of a complete set of reconfigurable photonic logic gates acting on XUV attosecond pulses and hence extends all-optical switching to the XUV and potentially the x-ray regime in the future.
While a readily realizable experiment on doubly-excited states in gas-phase helium is simulated in this work, attosecond FWM-based switching is a general concept.
With the main ingredients presented in this example---suitable electronic structure and laser parameters as well as leveraging strong-field effects---a clear pathway is presented to apply this FWM scheme to other targets and spectral regimes.
The background-free emission of the FWM signal will enable a single-shot XUV FWM measurement by replacing the CCD detector commonly used for measuring FWM spectograms with a motorized 2D slit and an XUV-sensitive photodiode.
Furthermore, heterodyning the FWM signal with an XUV reference that was split-off before the target cell, e.g. with a transmission-grating approach \cite{stooss2019xuv}, will grant access to the phase information of the XUV FWM emission.
This phase-sensitive detection together with extending the basis set to more Rydberg states leads the way to transforming this classical logic-gate approach into a quantum logic-gate implementation \cite{o2009photonic}.

\begin{backmatter}
\bmsection{Funding}
Alexander von Humboldt-Stiftung; Office of Science, Office of Basic Energy Sciences through the Atomic, Molecular, and Optical
Sciences Program of the Division of Chemical Sciences, Geosciences, and Biosciences of the U.S. Department of Energy
(DOE) at Lawrence Berkeley National Laboratory (DE-AC02-05CH11231).

\bmsection{Acknowledgment}
This work was performed by personnel supported by the Office of Science, Office of Basic Energy Sciences through the Atomic, Molecular, and Optical Sciences Program of the Division of Chemical Sciences, Geosciences, and Biosciences of the U.S. Department of Energy (DOE) at Lawrence Berkeley National Laboratory under Contract No. DE-AC02-05CH11231.
P.~Rupprecht thanks the Alexander von Humboldt Foundation for their support (Feodor Lynen Fellowship).
The authors thank D.~Cleveland for helpful discussions.

\bmsection{Disclosures}
The authors declare no conflicts of interest.

\bmsection{Data availability} Data underlying the results presented in this paper are not publicly available at this time but may be obtained from the authors upon reasonable request.

\bmsection{Supplementary material} See Supplement 1 for supporting content.

\end{backmatter}

\bibliography{sample}

\end{document}